\documentclass[12pt,usenatbib,referee]{mn2e}
\usepackage{psfig,natbib}

\def\la{\mathrel{\hbox{\rlap{\hbox{\lower4pt\hbox{$\sim$}}}\hbox{$<$}}}}
\def\ga{\mathrel{\hbox{\rlap{\hbox{\lower4pt\hbox{$\sim$}}}\hbox{$>$}}}}

\newcommand{\be}{\begin{eqnarray}}
\newcommand{\ee}{\end{eqnarray}}

\newcommand{\msol}{\ifmmode{{\rm M}_\odot}\else{M$_\odot$}\fi}
\newcommand{\foe}{\ifmmode{10^{51}}\else{$10^{51}$}\fi}

\newcommand{\xni}{\ifmmode{{\rm X}_{\rm Ni}}\else{X$_{\rm Ni}$}\fi}

\def\Teff{\ifmmode{T_{\rm eff}}\else{\hbox{$T_{\rm eff}$} }\fi}
\def\Rzero{\ifmmode{R_0}\else{\hbox{$R_0$} }\fi}

\def\SP2{{\tt IBM SP2}}

\def\PC2{{\tt PC$^2$}}

\def\logg{\log(g)}
\def\mh{[{\rm M/H}]}

\def\inu{\ifmmode{I_{\nu}}\else{\hbox{$I_{\nu}$} }\fi}
\def\snu{\ifmmode{S_{\nu}}\else{\hbox{$S_{\nu}$} }\fi}
\def\jnu{\ifmmode{J_{\nu}}\else{\hbox{$J_{\nu}$} }\fi}

\def\fep{\ifmmode{{\rm Fe II}}\else\hbox{Fe~II }\fi}

  at 12.0 true pt 

\def\b{\beta}

\def\rout{\ifmmode{r_{\rm out}}\else\hbox{$r_{\rm out}$}\fi}
\def\tmax{\ifmmode{\tau_{\rm max}}\else\hbox{$\tau_{\rm max}$}\fi}
\def\tstd{\ifmmode{\tau_{\rm std}}\else\hbox{$\tau_{\rm std}$}\fi}
\def\vmax{\ifmmode{v_{\rm max}}\else\hbox{$v_{\rm max}$}\fi}
\def\muE{\ifmmode{\mu_{\rm E}}\else\hbox{$\mu_{\rm E}$}\fi} 
\def\pE{\ifmmode{p_{\rm E}}\else\hbox{$p_{\rm E}$}\fi} 
\def\bmax{\ifmmode{\b_{\rm max}}\else\hbox{$\b_{\rm max}$}\fi}

\def\Teff{\hbox{$\,T_{\rm eff}$} }

\def\rout{\hbox{$r_{\rm out}$} }

\def\chistd{\ifmmode{\chi_{\rm std}}\else\hbox{$\chi_{\rm std}$}\fi}

\def\K{\,{\rm K}}
\def\msol{$M_\odot$}
\def\foe{10^{51}}

\def\xni{{\rm X}_{\rm Ni}}

\def\lstar{\ifmmode{\Lambda^*}\else\hbox{$\Lambda^*$}\fi} 
\def\Rop{\ifmmode{[R_{ij}]}\else\hbox{$[R_{ij}]$}\fi}

\def\Rji{\ifmmode{[R_{ji}]}\else\hbox{$[R_{ji}]$}\fi}
\def\Rstar{\ifmmode{[R_{ij}^*]}\else\hbox{$[R_{ij}^*]$}\fi}

\def\Rjistar{\ifmmode{[R_{ji}^*]}\else\hbox{$[R_{ji}^*]$}\fi}
\def\DRji{\ifmmode{[\Delta R_{ji}]}\else\hbox{$[\Delta R_{ji}]$}\fi}
\def\DRij{\ifmmode{[\Delta R_{ij}]}\else\hbox{$[\Delta R_{ij}]$}\fi}

\def\ns{\ifmmode{N_{\rm s}}          
        \else\hbox{$N_{\rm s}$}\fi}

\def\mat#1{{\bf #1}}     
\def\vek#1{{#1}}         

\newcount\eqcount
\def
  \nummer{
    \global\advance\eqcount by 1
    (\the\eqcount)
  }

\def
  \numadv{
    \global\advance\eqcount by 1
  }

\def
   \numout#1{
     (\the\eqcount #1)
  }

\def\ivek#1#2{\ifmmode{\vek{I}^{#1}_{#2}}
        \else\hbox{$\vek{I}^{#1}_{#2}$}\fi}

\def\tmat#1#2{\ifmmode{\mat{t}^{#1}_{#2}}
        \else\hbox{$\mat{t}^{#1}_{#2}$}\fi}
\def\rmat#1#2{\ifmmode{\mat{r}^{#1}_{#2}}
        \else\hbox{$\mat{r}^{#1}_{#2}$}\fi}
\def\bvek#1#2{\ifmmode{\beta^{#1}_{#2}}
        \else\hbox{$\beta^{#1}_{#2}$}\fi}

\def\bpi#1{\bvek{+}{#1}}

\def\lp{\ifmmode{\lambda^+_\tau}           
        \else\hbox{$\lambda^+_\tau$}\fi}
\def\lm{\ifmmode\lambda^-_\tau             
        \else\hbox{$\lambda^-_\tau$}\fi}

\chardef\tilt=126

\begin{document}

\newcommand\aj{{AJ}}%
\newcommand\araa{{ARA\&A}}%
\newcommand\apj{{ApJ}}%
\newcommand\apjl{{ApJ}}%
\newcommand\apjs{{ApJS}}%
\newcommand\ao{{Appl.~Opt.}}%
\newcommand\apss{{Ap\&SS}}%
\newcommand\aap{{A\&A}}%
\newcommand\aapr{{A\&A~Rev.}}%
\newcommand\aaps{{A\&AS}}%
\newcommand\azh{{AZh}}%
\newcommand\baas{{BAAS}}%
\newcommand\jrasc{{JRASC}}%
\newcommand\memras{{MmRAS}}%
\newcommand\mnras{{MNRAS}}%
\newcommand\pra{{Phys.~Rev.~A}}%
\newcommand\prb{{Phys.~Rev.~B}}%
\newcommand\prc{{Phys.~Rev.~C}}%
\newcommand\prd{{Phys.~Rev.~D}}%
\newcommand\pre{{Phys.~Rev.~E}}%
\newcommand\prl{{Phys.~Rev.~Lett.}}%
\newcommand\pasp{{PASP}}%
\newcommand\pasj{{PASJ}}%
\newcommand\qjras{{QJRAS}}%
\newcommand\skytel{{S\&T}}%
\newcommand\solphys{{Sol.~Phys.}}%
\newcommand\sovast{{Soviet~Ast.}}%
\newcommand\ssr{{Space~Sci.~Rev.}}%
\newcommand\zap{{ZAp}}%
\newcommand\nat{{Nature}}%
\newcommand\iaucirc{{IAU~Circ.}}%
\newcommand\aplett{{Astrophys.~Lett.}}%
\newcommand\apspr{{Astrophys.~Space~Phys.~Res.}}%
\newcommand\bain{{Bull.~Astron.~Inst.~Netherlands}}%
\newcommand\fcp{{Fund.~Cosmic~Phys.}}%
\newcommand\gca{{Geochim.~Cosmochim.~Acta}}%
\newcommand\grl{{Geophys.~Res.~Lett.}}%
\newcommand\jcp{{J.~Chem.~Phys.}}%
\newcommand\jgr{{J.~Geophys.~Res.}}%
\newcommand\jqsrt{{J.~Quant.~Spec.~Radiat.~Transf.}}%
\newcommand\memsai{{Mem.~Soc.~Astron.~Italiana}}%
\newcommand\nphysa{{Nucl.~Phys.~A}}%
\newcommand\physrep{{Phys.~Rep.}}%
\newcommand\physscr{{Phys.~Scr}}%
\newcommand\planss{{Planet.~Space~Sci.}}%
\newcommand\procspie{{Proc.~SPIE}}%
\let\astap=\aap 
\let\apjlett=\apjl 
\let\apjsupp=\apjs 
\let\applopt=\ao

\bibliographystyle{mn2e}

\title[Atmospheric Analysis of the M/L-- and M/T--Dwarf Binary Systems LHS~102 
and Gliese~229]{Atmospheric Analysis of the M/L-- and M/T--Dwarf Binary 
Systems LHS~102 and Gliese~229}

\author[S.K. Leggett et al]
{S.K. Leggett,$^1$
Peter H. Hauschildt,$^2$
F. Allard,$^3$ 
T.R. Geballe,$^4$
E.~Baron,$^5$ \\
$^1$Joint Astronomy Centre, University Park, Hilo, HI 96720,
Email: \tt skl@jach.hawaii.edu\\
$^2$Dept.\ of Physics and Astronomy \& Center for Simulational Physics, 
University of Georgia, Athens, GA 30602-2451;
Email: {\tt yeti@hal.physast.uga.edu}\\
$^3$CRAL, Ecole Normale Superieure, 46 Allee d'Italie, Lyon, 69364 France;
Email: {\tt fallard@cral.ens-lyon.fr } \\
$^4$Gemini Observatory, University Park, Hilo, HI 96720;
Email: {\tt tgeballe@gemini.edu}\\
$^5$Dept. of Physics and Astronomy, University of
Oklahoma, 440 W.  Brooks, Rm 131, Norman, OK 73019-0260;
Email: {\tt baron@nhn.ou.edu}
}

\maketitle
\begin{abstract}

We present  0.9---2.5$\mu$m spectroscopy with  R$\sim$800
and 1.12-1.22$\mu$m spectroscopy with  R$\sim$5800 for the 
M dwarfs Gl 229A and LHS 102A, and for the L dwarf LHS 102B.  
We also report $IZJHKL^{\prime}$ photometry for both components of 
the LHS~102 system, and  $L^{\prime}$ photometry for Gl~229A.
The data are combined with previously published spectroscopy
and photometry to produce flux distributions for each component
of the kinematically old disk M/L--dwarf binary system LHS~102
and the kinematically young disk M/T--dwarf binary system Gliese~229.
The data are analyzed using synthetic spectra generated by the latest 
``AMES-dusty'' and ``AMES-cond'' models by Allard \& Hauschildt.  Although 
the models are not able to reproduce the overall slope of the 
infrared flux distribution of the L dwarf, most likely due to the treatment 
of dust in the photosphere, the data for the M dwarfs and the T dwarf are 
well matched.  We find that the Gl~229 system is metal--poor despite having
kinematics of the young disk, and that the LHS~102 system has solar
metallicity.   The observed luminosities and derived temperatures 
and gravities are consistent with evolutionary model predictions if the
Gl~229 system is very young (age $\sim 30\,$Myr) with masses (A,B) of
(0.38,$\ga$ 0.007)M$_{\sun}$, and the LHS~102 system is older,
aged 1---10~Gyr with masses (A,B) of (0.19,0.07)M$_{\sun}$.

\end{abstract}

\section{Introduction}

In the last few years there have been dramatic developments in the study of
very low mass stars and brown dwarfs --- where the latter are defined to be
objects with mass below that required for stable hydrogen burning.
Sky surveys have found a significant population of low mass dwarfs
cooler than M--dwarfs.  The first population to be identified was the  L--dwarf
class, distinguished from the M--dwarfs by weakening VO and TiO absorption features
in the red, and by stronger alkaline and H$_2$O absorption features in
the red and near--infrared  \cite[e.g.][]{d97,k99,mar99}.  The L--dwarfs
cover an effective temperature range of about 2200~K to 1400~K 
\cite[e.g.][]{l01} and the prototype of this class is
GD~165B, discovered by \cite{bz88} as a red companion to
a hot white dwarf.  Several objects even cooler than the L--dwarfs
were discovered in 1999 by the Sloan Digital Sky Survey and the 2 Micron 
All--Sky Survey \cite[]{str99,bur99}.  They are very similar to Gliese 229B, 
the extremely low--mass companion to an M--dwarf found by
\cite{nak95}, and are distinguished by the presence of CH$_4$ absorption
in the near--infrared H and K bands.  These objects, known as T--dwarfs, 
have $\Teff \sim$1300---800~K and tens of T--dwarfs are now known
\cite[see][and references therein]{g01,b01}.

In this paper we present an analysis of the low--mass binary
systems LHS~102 and Gliese (Gl)~229.  The primaries of these two
systems are  M--dwarfs which are in the proper motion catalogues by
\cite{luy79} and \cite{gj91}.  The secondaries were only recently detected.
Gl~229B, the prototypical T--dwarf, was discussed above;
LHS~102B is an L dwarf discovered by \cite{go99}.  We present new  UKIRT
infrared photometry and spectroscopy of these objects in \S 2, which
are analyzed using the latest models by Allard \& Hauschildt
\citep*{ng-hot}, described in \S 3.  The results of the model comparison
are given in \S 4, the implications for age and mass of each component are
discussed in \S 5 and our conclusions given in \S 6.
These two systems span the newly defined low--mass spectral classes and,
being binaries at known distances where each component has presumably the 
same chemical composition and age, are potentially very useful for constraining 
atmospheric and evolutionary models of ultracool dwarfs.

\section{Observational Results}

\subsection{The Sample}

Table 1 gives the LHS and Gliese or Gliese--Jahreiss catalogue numbers
\cite[]{luy79,gj91} for the two systems discussed in this work.  Although
both primaries are in both catalogues we  follow tradition and refer
to one as LHS~102 and the other as Gl~229.  The coordinates 
in the Table are based on our observations at UKIRT and are accurate
to about an arcsecond.  

Spectral types also are listed in Table 1.   For the M dwarfs LHS~102A 
and Gl~229A  these have been taken from \cite{mar99} and \cite{l92}, 
respectively.  The T subtype for Gl~229B is taken from the provisional 
classification scheme of \cite{g01}.  For LHS~102B the indices
presented by Geballe et al. imply  L4.0 based on the red spectrum,
L4.0 based on the K spectrum, and a later type of L6.5 based on the 
H--band water index.   As the last region is slightly noisier than
the others we adopt a type of L4.5$\pm$0.5.  This is consistent with
previously published classifications based on red spectra only: \cite{mar99} 
classify this object as L4, \cite{k00} classify it L5.  

Table 1 also gives the kinematic population for each system, based on
the space motion of the primaries, where distances were obtained from the
Hipparcos Catalogue for  Gl~229 \cite[]{per97} and from the Yale Catalogue 
for LHS~102 \cite[]{van94}.  The population for the Gl~229 system is from
\cite{l92}; we determine the population for the LHS~102 system to be older, 
based on the specifications given by Leggett for UVW space motions.  The 
space motions for LHS~102A were calculated
using a radial velocity from \cite{ro74}; a more recent
but quite different value for the secondary from \cite{bas00} gives the
same end result --- that the V motion at  $-73$ (for LHS~102A) or 
$-77$km/s (for LHS~102B) is too large for the ``young'' disk.

\subsection{New Photometry and Observed Colours}

New photometry has been obtained for three of the stars in the sample.
In 1999 August, $IZJHK$ was obtained for the LHS~102 system using 
the UK Infrared Telescope (UKIRT) with the
UFTI camera, a 1024$\times$1024 imager with 0$\farcs$09 pixels.  The
$I$ and $Z$ filters are non--standard and have
half--power bandwidths of  0.72---0.93~$\mu$m, and 0.85---1.055~$\mu$m.
\cite{lan92}  standards were used to calibrate  the $I$ data and provisional
Sloan standards were used for the $Z$ data \cite[]{kris98}. For both $I$ and
$Z$ there are strong colour terms between the UFTI system and the Landolt
and Sloan systems; the results are given in Table 2 in the natural UFTI
system.  

The UFTI $JHK$ filters are part of Mauna Kea consortium filter set
\cite[]{si01,to01}.  UKIRT faint standards \cite[]{haw01} were used to 
calibrate the photometry.  There are colour differences between the IRCAM3
system used to establish the UKIRT faint standards and the natural
UFTI system, defined by the new filter set \cite[for more discussion see][]
{haw01,l01}.  In Table 2 we list $JHK$ in the 
natural UFTI or Mauna Kea consortium (also known as MKO) system.

New L$^{\prime}$ photometry was obtained for LHS~102 A and B, and Gl~229A,
in 1999 September, using UKIRT's IRCAM/TUFTI camera, a 256$\times$256
imager with 0$\farcs$08 pixels.  Bright UKIRT standards were used to
calibrate the photometry.  The L$^{\prime}$ magnitudes of late L dwarfs and 
T dwarfs are different for  the Mauna Kea consortium filter and the filter
previously installed in IRCAM and referred to as the UKIRT filter
\cite[see][]{l01,ds01}; the results in Table 2 are on the MKO system.

Table 3 gives for both systems the distances taken from the parallax catalogues
\cite[]{per97,van94} in the form of distance moduli, and tabulates
$IZJHKL^{\prime}$ colours for the sample.  In this Table we give $I$ magnitudes
on the Cousins system.  The value for Gl~229A  is taken from the compilation by 
\cite{l92}. For the other three objects the $I_C$ value
is synthesized from the flux--calibrated spectral energy distributions
using a filter profile from \cite{bes90} (red spectra from the literature have 
been added to the infrared spectra, see \S 2.3).  For 
Gl~229B we extended the available spectroscopic data shortwards
of 0.8$\mu$m to 0.7$\mu$m using as a template the energy distributions for
the similar object in \cite{str99}.  The uncertainties in $I$ for Gl~229B 
and LHS~102A,B are $\sim 10$\%.  We suggest that these values for $I$ are 
preferable to those reported by \cite{go99} for LHS~102 A and B, which are 
inconsistent with our flux distributions, as is the value for LHS~102A reported 
by \cite{ro74}; i.e. adoption of either of these earlier results would lead to 
a discontinuity in the observed red to near--infrared flux distribution.
The $Z$ magnitudes for Gl~229 A and B are synthesized from their spectral energy 
distributions (the primary would saturate UFTI and the secondary is contaminated 
by the primary in imaging mode); the uncertainties are $\sim 10$\%.   The MKO 
system $JHK$ for Gl~229A are synthesized from our flux--calibrated spectral energy 
distributions \cite[the spectra were flux calibrated using the photometry reported 
in][]{l92}; the uncertainties are $\sim 5$\%.

\subsection{New Spectroscopy and Spectroscopic Sequences}

New R$\sim$5800 spectroscopic data for 1.12---1.22$\mu$m for Gl~229A and 
LHS~102A,B were obtained during 1999 August---November, using the CGS4 spectrometer 
on UKIRT. Telluric features were removed by ratioing with a nearby A-- or F--type 
star (after removing hydrogen lines present in the ratio star 
spectrum).  The data were obtained in two spectral segments which were then
joined around 1.16--1.17~$\mu$m.  Data with R$\sim$3000 over narrow intervals 
in the $J$, $H$ and $K$ bands for Gl~229B were taken from Saumon et al. (2000)
and also used in this analysis.

Spectra at R$\sim$800 covering 0.86---2.52~$\mu$m for Gl~229A and LHS~102A, and 
0.83---2.52~$\mu$m for LHS~102B, were obtained on 1999 September and August, 
also using CGS4  on UKIRT.  Telluric features were removed by ratioing with nearby
A-- or F--type stars (after removing hydrogen lines present in the ratio star 
spectrum).  The spectra consisted of several segments, each segment was
flux calibrated using our photometry and conjoined with adjacent segments to 
form spectral energy distributions.  We excluded  the range 1.36---1.41~$\mu$m 
where terrestrial H$_2$O absorption makes the data too noisy to be useful.  

The 0.84---4.15~$\mu$m spectra for Gl~229B is taken from \cite{l99} who 
re--flux--calibrated the red and L--band spectra of \cite{opp98} and the 
near--infrared spectra of \cite{ge96} using  photometry.  These data  have 
been further refined by putting the Oppenheimer et al. data onto vacuum wavelengths 
and calibrating them using the HST photometry of \cite{gol98}.  We also use the 
R$\sim$480  4.5---5.1~$\mu$m  spectrum of Gl~229B from \cite{noll97}. This 
spectrum is not flux calibrated photometrically, and the flux level may be 
uncertain by $\sim$30\% (due to possible differences in slit losses between 
Gl~229B and the calibration star).

Spectra shortwards of 0.86~$\mu$m have been taken from \cite{h94} for
Gl~229A (R$\approx$400), from \cite{he00} for LHS~102A  (R$\approx$700),
and from \cite{mar99} for LHS~102B (R$\approx$700).  The blue ends of the CGS4
spectra were calibrated using our $Z$ and $J$ photometry and the red spectra were
scaled to match the CGS4 data in the region of overlap.

\subsection{Integrated Fluxes and Bolometric Corrections}

Table 3 gives integrated fluxes for the sample, expressed as fluxes at the
Earth, bolometric magnitudes and intrinsic stellar luminosities. The updated
results for Gl~229B are expressed on the MKO system \cite[also presented 
in][]{l01}.  For the remaining objects, the integrated fluxes were
obtained by integrating the spectroscopic data over wavelength, and adding the
flux contributions at shorter and longer wavelengths.
The shorter wavelength flux contribution was adopted to be  a
simple linear extrapolation to zero flux at zero wavelength for LHS~102B,
but for Gl~229A the B magnitude \cite[]{l92}, and for LHS~102A the
B and V magnitudes \cite[]{ro74}, were used to estimate the shorter wavelength
flux distributions, using an effective wavelength approach.
The flux contributions at  wavelengths beyond 2.4$\mu$m were
calculated by deriving the fluxes at L$^{\prime}$ using an effective wavelength
approach, including the contributions between the long wavelength end of the 
K--band spectrum to the L$^{\prime}$ wavelength, determined by
linear interpolation, and assuming Rayleigh--Jeans tails beyond L$^{\prime}$. 
The last is a reasonable assumption for M-- and mid--L--dwarfs, but not for 
methane dwarfs, as discussed in \cite{l99}.  The uncertainties are 5\%, 0.05 mag 
and 0.02dex in total flux,  bolometric correction and log$_{10}L/L_{\odot}$, 
respectively.  The luminosities are used later in our analysis, together with 
the derived atmospheric parameters, to constrain masses and ages for the binaries.

\section{Models and Comparison to Synthetic Spectra}

\subsection{Models}

The  models  used for this  work were calculated as
described in  \cite{LimDust}, and are based on the Ames H$_2$O  and TiO line
lists by  \cite{ames-water-new} and \cite{ames-tio}.  We  stress  that large
uncertainties persist  in these opacities for the  temperature range of this
work \cite[see][]{2000ApJ...539..366A}.  Test calculations performed by 
replacing the Ames water  with the water line list from 
the SCAN database \cite[]{scan-h2o-new}, for somewhat lower temperatures, show  
relatively small effects at low spectral resolution; therefore we retained
the Ames H$_2$O line list for this analysis.  The models  and their comparisons 
to earlier versions  are the subject of a  separate publication \cite[]{LimDust} 
and we thus do not repeat their detailed description here. 

The models  have been upgraded with (i) the  replacement  of  the  JOLA  
(Just Overlapping  Line Approximation) opacities for FeH, VO  and CrH by 
opacity sampling line data  from \cite{FeHberk2}  and  R.  Freedman  (NASA-Ames,  
private communication), and (ii)  the extension of our database  of dust grain 
opacities from 10  to 40 species.  
In the following, these models will be referred to as ``AMES-dusty'' for 
models in which the dust particles stay in the layers in which they have
formed, and ``AMES-cond'' for models in which the dust particles have 
sunk below the atmospheric layer where they originally formed (and below the 
photosphere).  The term ``AMES grid'' refers to either of the ``AMES-dusty'' 
or  ``AMES-cond'' models.

In order to investigate the effects of mixing length on the model structure 
and synthetic spectra, the AMES grid was extended to include mixing lengths 
$\ell$ of $1.5$ and $2.0$ times the local pressure scale height $H_p$ for the 
parameter range of interest. These additional models were otherwise constructed 
with the same setup as those discussed in \cite{LimDust} to avoid systematic 
errors in the analysis.

The primaries (A components) analyzed in this paper have effective temperatures 
that are too high for dust to form and to affect their atmospheres. 
These relatively high effective temperatures allow us to compare the data with 
the NextGen grid of model atmospheres for dwarfs \citep*{ng-hot} and
giants \cite[]{ng-giants} and thus to compare the results
obtained using older and newer model grids. The T dwarf Gl 229B is in the regime 
where dust has formed and settled below the atmosphere \cite[cf.][]{LimDust} 
whereas the L dwarf LHS 102B is in the regime where dust formation and opacities 
have to be considered. For each of these cooler objects we use the newer and 
appropriate models for the analysis (see \S 4.2 and \S 4.4).

\subsection{Comparison Process}

Several model atmosphere grids were generated, covering the  range 
$500\K\leq \Teff \leq 4000\K$, $3.5\leq \logg \leq 5.5$ and 
[M/H]$= -1.0, -0.7, -0.5$ and 0.0 for mixing lengths 
of 1.0, 1.5 and 2.0, for  a total  of  $\approx 300$  model  atmospheres. 
Synthetic spectra generated by these models were compared to the observed spectra  
using an IDL program.    In addition, the
observed spectra were compared against the NextGen grid \citep*{ng-hot} over a
wider range in metallicity but for only one value of the mixing length, in order 
to assess the stability of the derived parameters and to search for systematic
problems.  

First, the resolution of the synthetic spectra was
degraded to  that  of  each  observed  spectrum by convolution with
a Gaussian of the appropriate width, and  the spectra were normalized  to unit
area for scaling.   Next, for  each observed spectrum  the program calculated  a 
quality function $q$,  similar to a $\chi^2$ value,  for the comparison  with all 
synthetic spectra  in the grid. The quality function is calculated by first
scaling the model spectrum to the observed fluxes, then mapping 
the synthetic spectrum (reduced to the resolution of the observed data) onto
the grid of observed wavelength points, and, finally, calculating
$$
q = \sum_i w_i \left(0.5\frac{f^{\rm model}_i-f^{\rm obs}_i}{f^{\rm model}_i+f^{\rm obs}_i}\right)^2
$$
with $w_i = 0.5(f^{\rm obs}_{i+1}+f^{\rm obs}_{i})(\lambda^{\rm obs}_{i+1}-\lambda^{\rm obs}_{i})$
where $f^{\rm model}$ is the (mapped) flux of the model spectrum,
$f^{\rm obs}$ is the observed flux, and $\lambda^{\rm obs}$ the observed 
wavelength.

We selected the models that resulted in 
the 3---10 lowest $q$ values as the most probable  parameter range  for  each 
individual  star.  The ``best'' value was chosen by  visual inspection. This 
procedure  allows a
rough estimate of the uncertainty in the  stellar parameters. Note that it does
not eliminate systematic errors in the stellar parameters due to missing,
incorrect or incomplete opacity sources.  Table 4 lists model parameters and
the goodness of fit value for each object, for models with a range of
parameters selected to cover the probable  range.  Only the relative
value of  $q$ is important, the actual value is arbitrary as it depends on the 
number of data points used and on the absolute flux levels; a decreasing value 
indicates a better fit.

\section{Results}

\subsection{LHS 102A}

\begin{figure}
\psfig{file=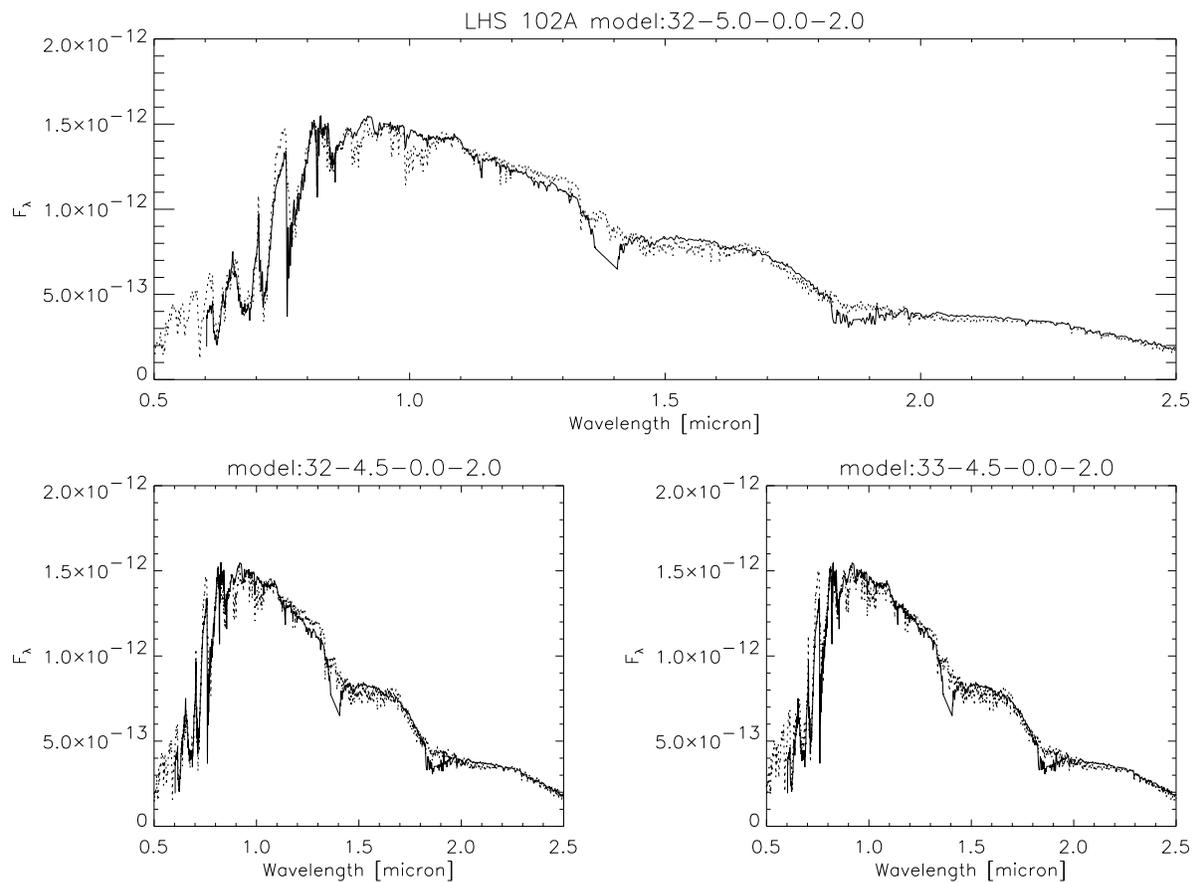,width=0.99\hsize,angle=90}
\caption[]{\label{lhs102a}Best fit to LHS102A using our current model
grids. The parameters of the models (dotted lines) are 
$\Teff=3200\K$, $\logg=5.0$, $\alpha=2.0$ (top panel);
$\Teff=3200\K$, $\logg=4.5$, $\alpha=2.0$ (bottom left panel);
$\Teff=3300\K$, $\logg=4.5$, $\alpha=2.0$ (bottom right panel).
All models shown have solar abundances.
}
\end{figure}

The spectrum of LHS 102A together with the best fitting synthetic spectra 
generated from the Ames grid models is presented in Fig.\ \ref{lhs102a}.
The best fitting models have  $\Teff\approx 3200\K$ and $\logg\approx 5.0$,
with solar abundances. The formal uncertainty is about $100\K$ in $\Teff$ and 
$0.5$ in $\logg$.  Metal poor models gave a significantly worse fit
as indicated in Table 4.  NextGen model fits to LHS 102A result in derived 
parameters of $\Teff \approx 3400\K$, $\logg \approx 5.5$ and solar abundances. 
However although the NextGen models include both metal-rich
and metal-poor models, the mixing length  is
fixed at $\alpha = 1.0$. Considering the results from all sets of fits, we
conclude that $\Teff\approx 3200\K $, $\logg \approx 5.0$, solar
abundances, and $\alpha = 2.0$ are the likely parameters for LHS 102A.

\begin{figure}
\psfig{file=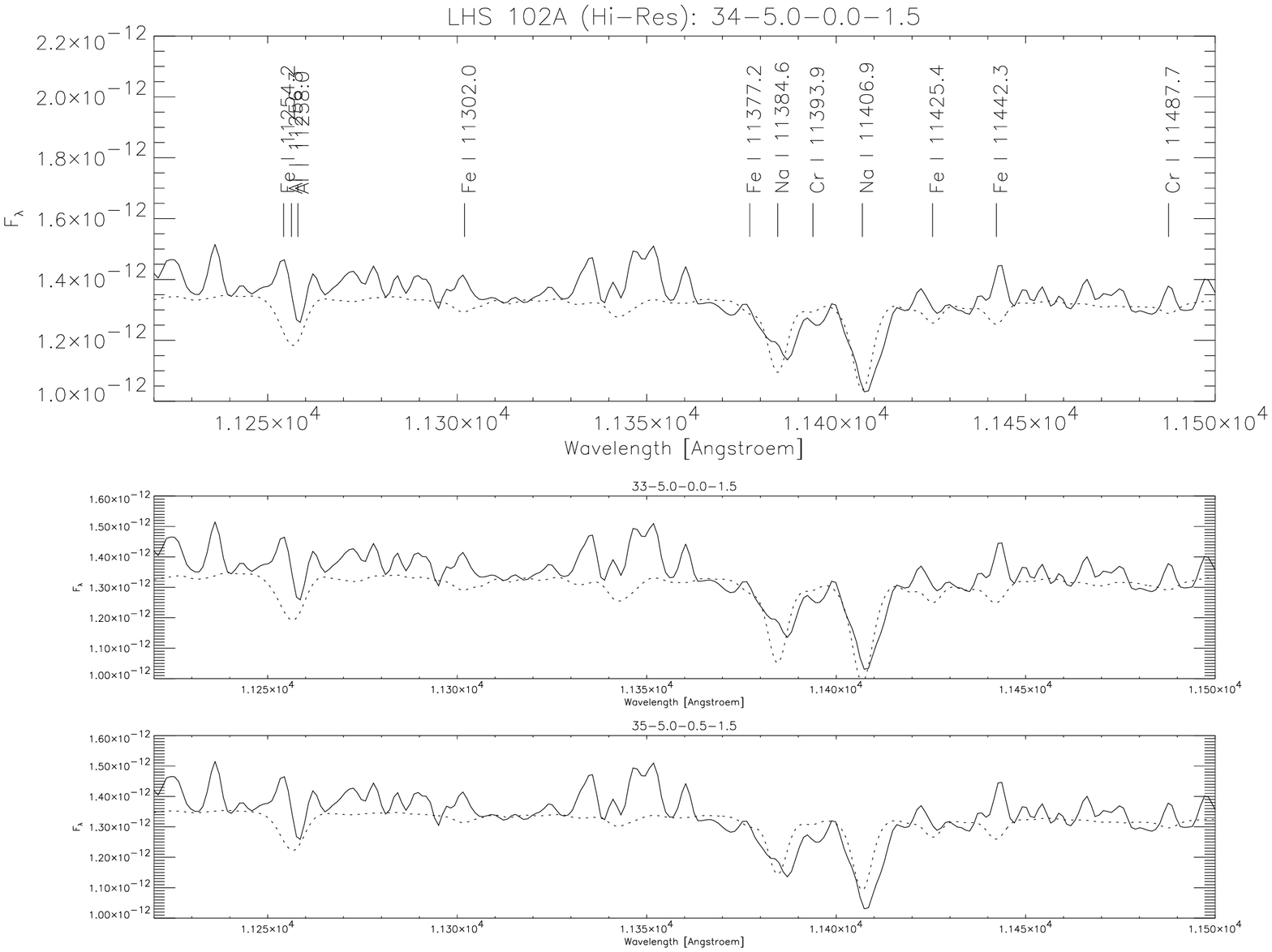,width=0.99\hsize,angle=90}
\caption[]{\label{lhs102a-hr-short}Fit to higher resolution spectrum of LHS 102A using
best fit high-resolution spectrum plus a number of close-by models. The parameters of the models (dotted lines) are 
$\Teff=3400\K$, $\logg=5.0$, $\mh=0.0$ (top panel);
$\Teff=3300\K$, $\logg=5.0$, $\mh=0.0$ (middle panel);
$\Teff=3500\K$, $\logg=5.0$, $\mh=-0.5$ (bottom panel).
All models shown have mixing lengths $\alpha=1.5$.}
\end{figure}

\begin{figure}
\psfig{file=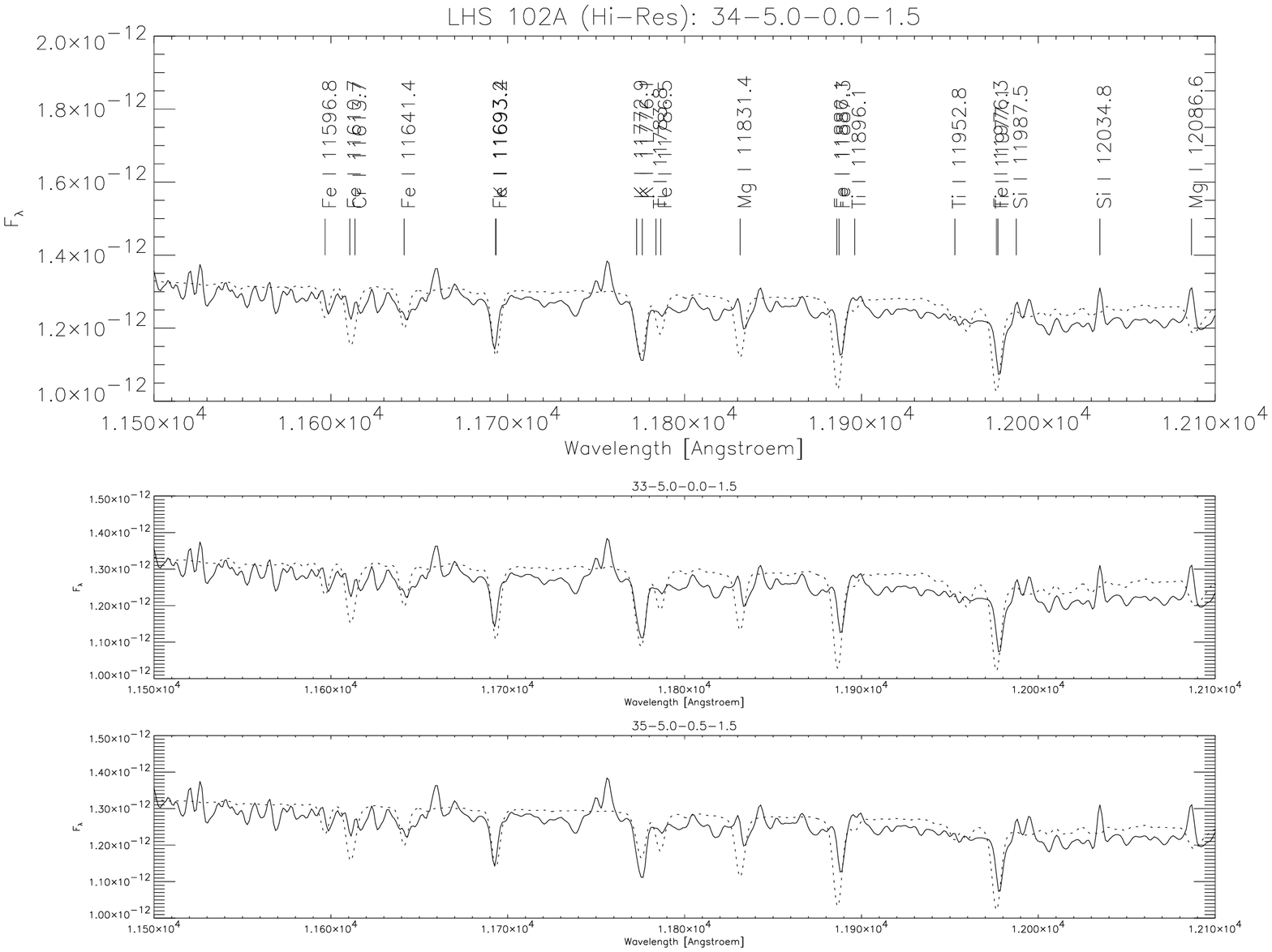,width=0.99\hsize,angle=90}
\caption[]{\label{lhs102a-hr-long}Fit to higher resolution spectrum of LHS 102A using
best fit high-resolution spectrum plus a number of close-by models. The parameters of the models (dotted lines) are 
$\Teff=3400\K$, $\logg=5.0$, $\mh=0.0$ (top panel);
$\Teff=3300\K$, $\logg=5.0$, $\mh=0.0$ (middle panel);
$\Teff=3500\K$, $\logg=5.0$, $\mh=-0.5$ (bottom panel).
All models shown have mixing lengths $\alpha=1.5$.}
\end{figure}

The fits to the high-resolution spectrum of LHS 102A are shown in
Figs.~\ref{lhs102a-hr-short} and \ref{lhs102a-hr-long}.  We have used the 
Ames grid models to generate high-resolution synthetic spectra in the parameter 
range $(\Teff,\logg)$ found from the low-resolution fitting and selected
the three models that best fit  the high-resolution spectra. Many of the 
observed lines are reproduced quite well in the synthetic spectra, indicating 
that the derived parameters and their ranges are good. There are no strong 
indications for non-solar patterns that can be established with confidence 
from the data, but spectral lines from more elements are required to confirm 
this result. 

\subsection{LHS 102B}

\begin{figure}
\psfig{file=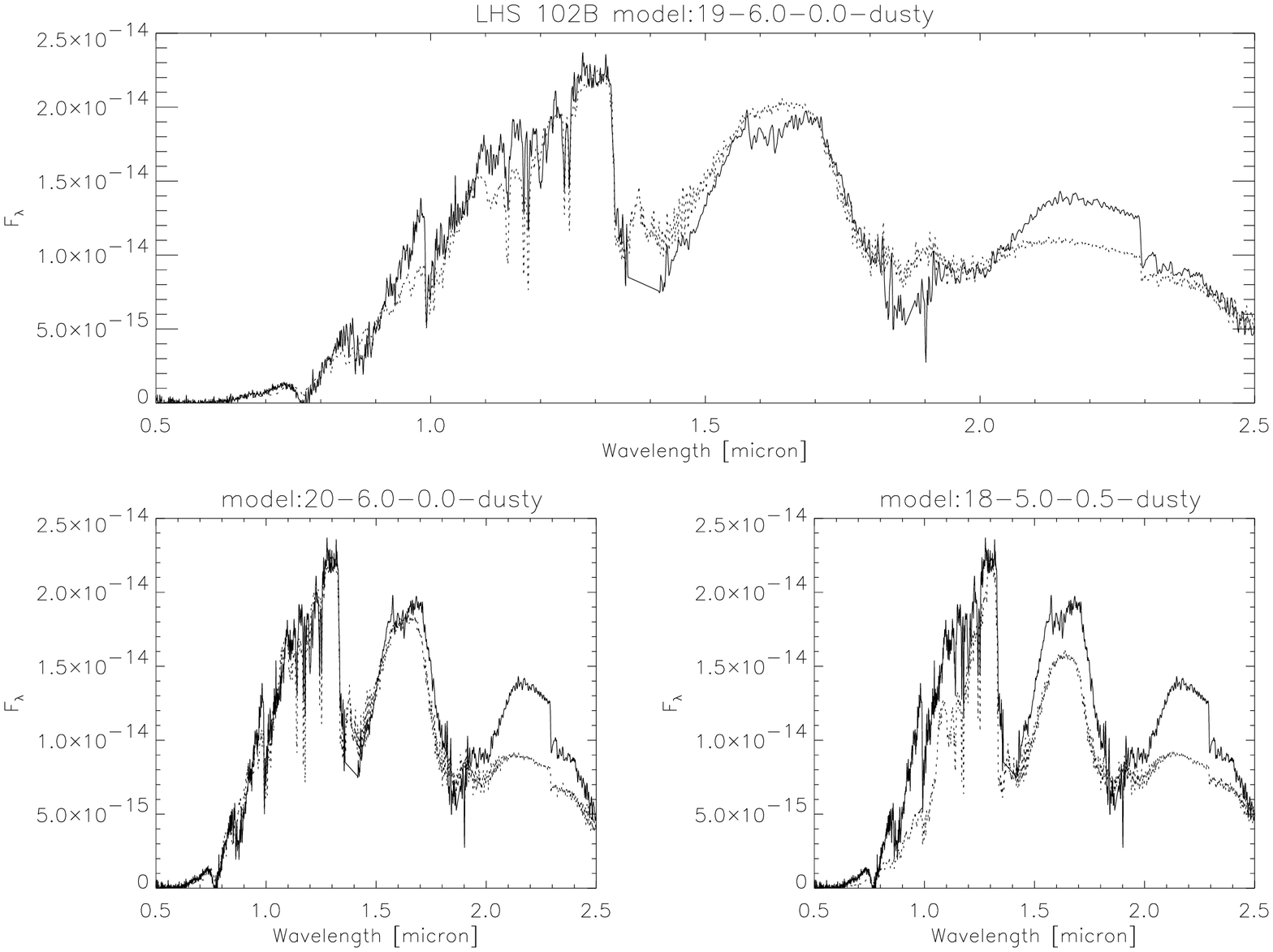,width=0.99\hsize,angle=90}
\caption[]{\label{lhs102b}Best fits to LHS102B using ``AMES-dusty'' models.
The parameters of the models (dotted lines) are 
$\Teff=1900\K$, $\logg=6.0$, $\mh=0.0$ (top panel);
$\Teff=2000\K$, $\logg=6.0$, $\mh=0.0$ (bottom left panel);
$\Teff=1800\K$, $\logg=5.0$, $\mh=-0.5$ (bottom right panel).
}
\end{figure}

\begin{figure}
\psfig{file=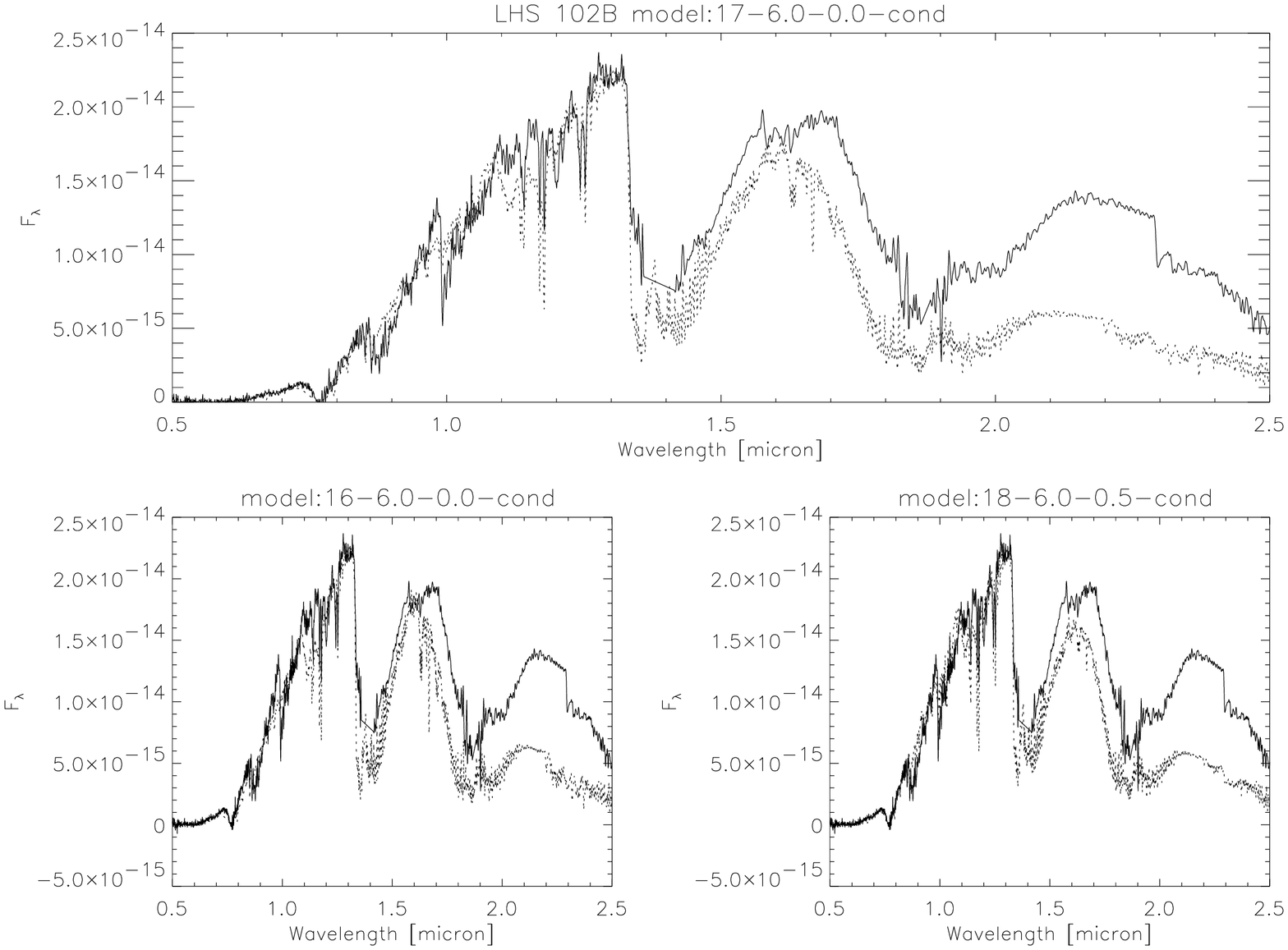,width=0.99\hsize,angle=90}
\caption[]{\label{lhs102b-cond}Best fits to LHS102B using ``AMES-cond'' models.
The parameters of the models (dotted lines) are 
$\Teff=1700\K$, $\logg=6.0$, $\mh=0.0$ (top panel);
$\Teff=1600\K$, $\logg=6.0$, $\mh=0.0$ (bottom left panel);
$\Teff=1800\K$, $\logg=5.0$, $\mh=-0.5$ (bottom right panel).
}
\end{figure}

In Fig.~\ref{lhs102b} we show the observed spectrum and best fits to LHS 102B 
using ``AMES-dusty'' models from \cite{LimDust}; for comparison we show the results 
obtained using ``AMES-cond'' models (in which the dust grains are assumed to have 
settled completely below the atmosphere) in Fig.~\ref{lhs102b-cond}. As to be expected, 
the ``AMES-dusty'' models result in substantially better fits to the water bands than 
the ``AMES-cond'' models, although neither reproduces the heights of both the J--band 
and K--band flux peaks.   LHS 102B's derived atmospheric parameters are 
$\Teff=1900\K\pm 100\K$, $\logg \approx 6.0$, solar abundances; at this temperature 
dust formation and opacities are important \cite[]{LimDust}. The parameters derived by 
the fitting procedure using ``AMES-dusty'' and ``AMES-cond'' models are quite similar.  
This increases our confidence in the derived set of parameters, as the ``AMES-dusty'' 
and ``AMES-cond'' models are extreme cases and partial settling models (under 
construction) will fall in between these limits \cite[see][for more
details]{LimDust}.  The model parameters listed in Table 4 are for the ``AMES-dusty''
model grid.

\begin{figure}
\psfig{file=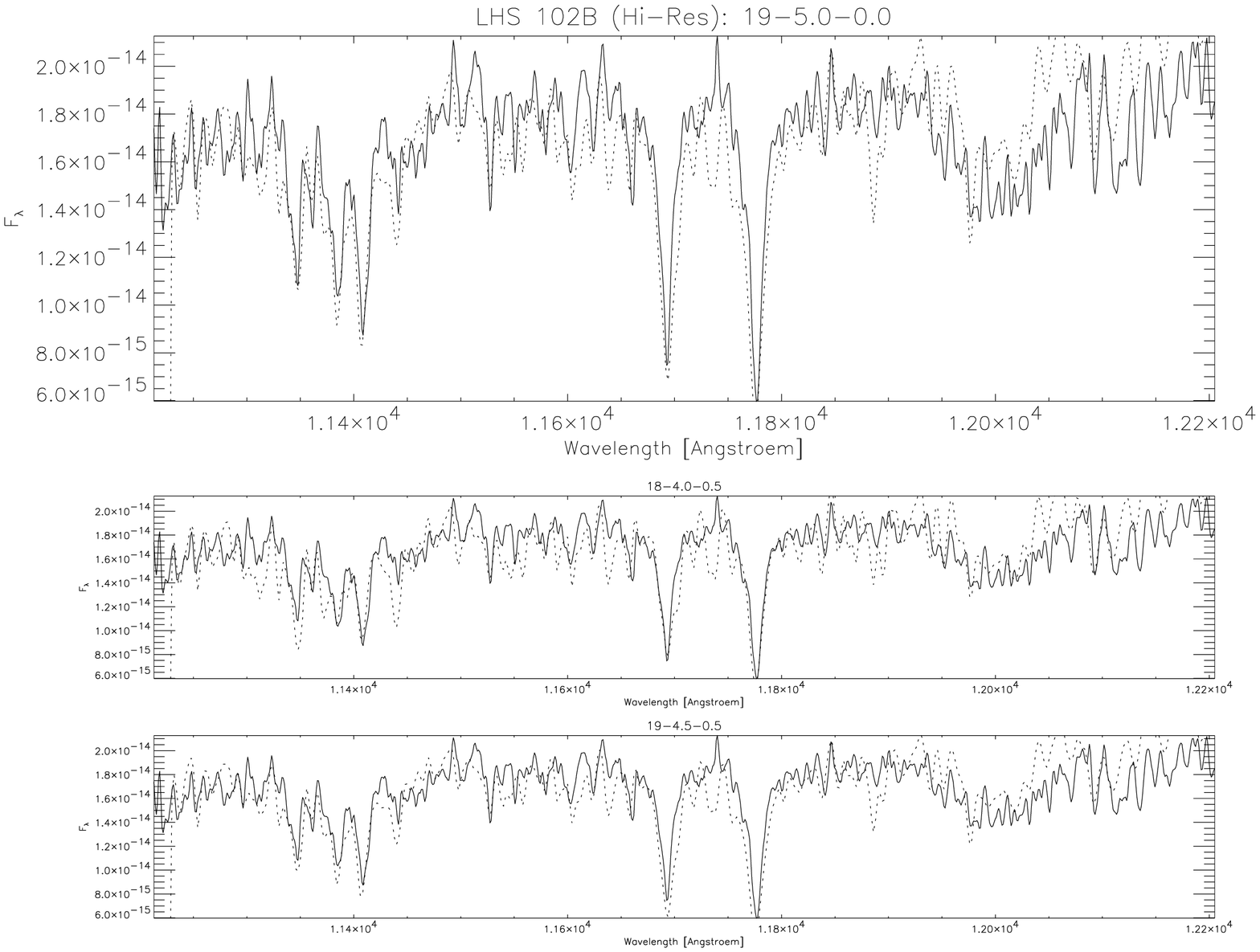,width=0.99\hsize,angle=90}
\caption[]{\label{lhs102b-hr}Fit to higher resolution spectrum of LHS 102B using
best fit high-resolution spectrum plus a number of close-by models. The parameters of the models (dotted lines) are 
$\Teff=1900\K$, $\logg=5.0$, $\mh = 0.0$ (top panel);
$\Teff=1800\K$, $\logg=4.0$, $\mh = -0.5$ (middle panel);
$\Teff=1900\K$, $\logg=4.5$, $\mh = -0.5$ (bottom panel).
}
\end{figure}

The ``AMES-dusty'' fit to the high resolution spectrum of LHS 102B is shown in 
Fig.~\ref{lhs102b-hr}.  The synthetic spectra fit the atomic line profiles quite well 
and many molecular bands (hundreds of overlapping line for multiple systems) are 
reproduced reasonably well. As in the case of LHS 102A, the spectrum is well fit  
with solar abundances.

\subsection{Gl 229A}

\begin{figure}
\psfig{file=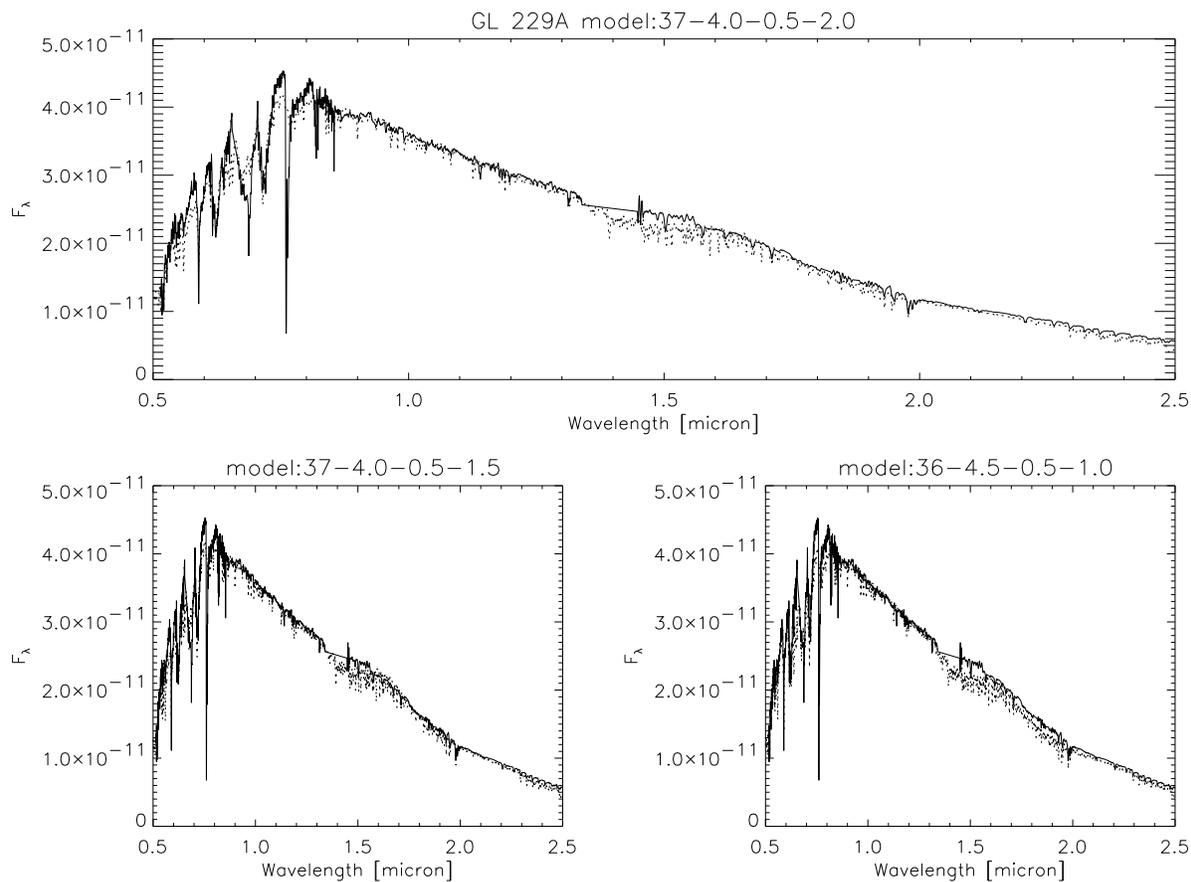,width=0.99\hsize,angle=90}
\caption[]{\label{gl229a}Best fit to Gl229A using the newest model grids.
The parameters of the models (dotted lines) are 
$\Teff=3700\K$, $\logg=4.0$, $\mh=-0.5$, $\alpha=2.0$ (top panel);
$\Teff=3700\K$, $\logg=4.0$, $\mh=-0.5$, $\alpha=1.5$ (bottom left panel);
$\Teff=3600\K$, $\logg=4.5$, $\mh=-0.5$, $\alpha=1.0$ (bottom right panel).
}
\end{figure}

Fig.~\ref{gl229a} shows the best fit to the low resolution spectrum of Gl 229A.
The automatic fitting procedure yielded a best fitting model from
the Ames grid with the parameters $\Teff=3700\K$, $\logg=4.0$, $\mh=-0.5$
and a mixing length parameter $\alpha =\ell/H_p = 2.0$.
 However, a model with $\alpha=1.5$ fits only marginally
worse (Fig.~\ref{gl229a}) for the same effective temperature and
gravity. The third best fit is significantly worse in quality than the first two
and gives parameters of $\Teff=3600\K$, $\logg=4.5$, $\mh=-0.5$ and $\alpha =
1.0$.  Table 4 shows that solar metallicity models gave significantly worse fits.
Repeating the fitting procedure with NextGen
models gave best fits in the range of $3500\K \le \Teff \le 3600\K$, $3.5 \le
\logg \le 4.0$, $-1.5 \le \mh \le -1.0$.  The mixing length of the NextGen
grid is fixed at $\alpha = 1.0$ and the quality of the NextGen
fits is noticeably worse, the AMES grid models (even at $\alpha = 1.0$) fit the
observed spectra much better. We conclude that Gl 229A is somewhat metal
poor, $\mh \approx -0.5$, and that the best fitting mixing length parameter
$\alpha$ is between $1.5$ and $2.0$. The latter is consistent with 3D
hydrodynamical models of objects in this effective temperature regime
\cite[]{ludwig-pap}. 

\begin{figure}
\psfig{file=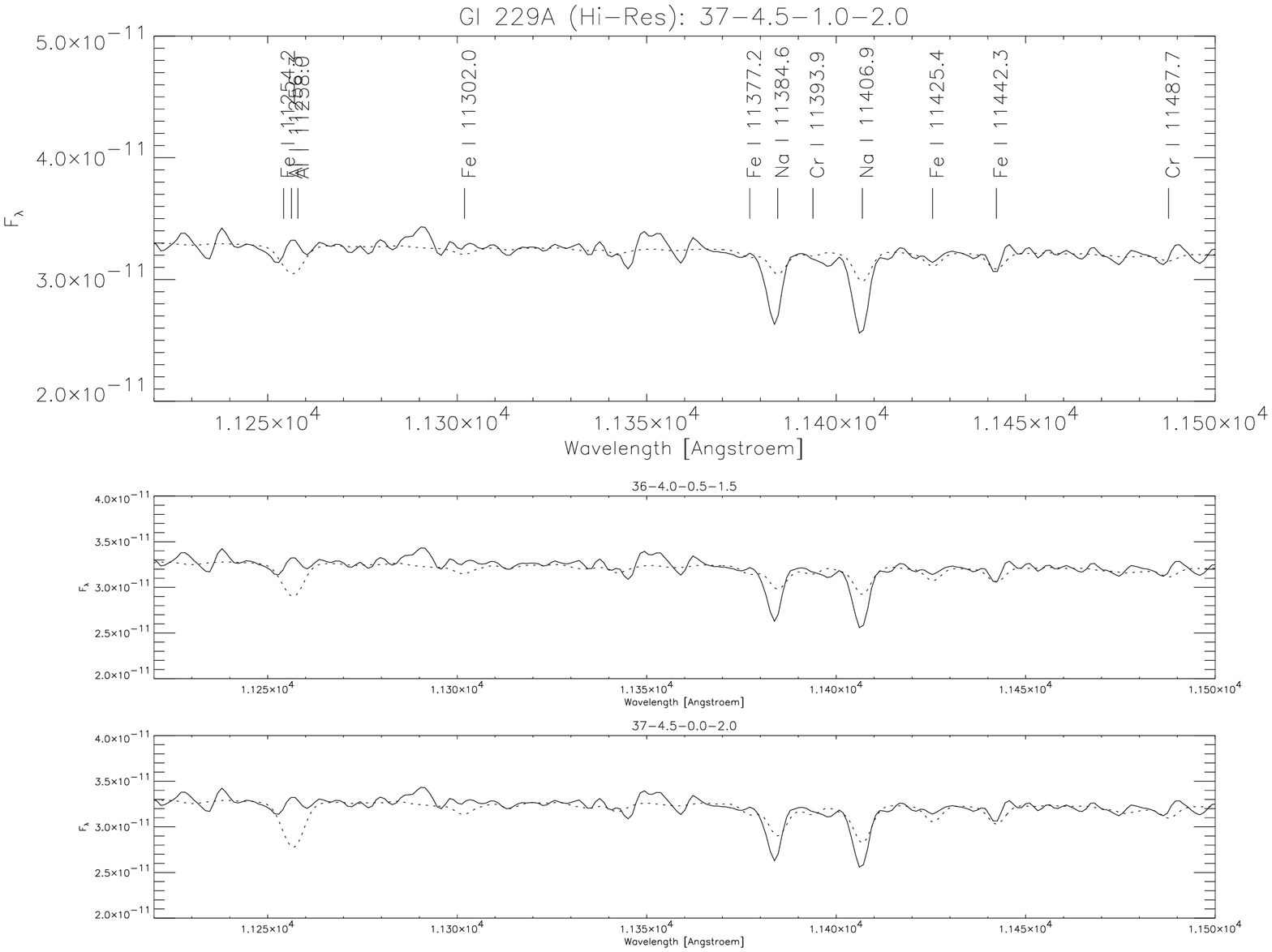,width=0.99\hsize,angle=90}
\caption[]{\label{gl229a-hr-short}Fit to higher resolution spectrum of Gl 229A using
best fit high-resolution spectrum plus a number of close-by models. The parameters of the models (dotted lines) are 
$\Teff=3700\K$, $\logg=4.5$, $\mh=-1.0$, $\alpha=2.0$ (top panel);
$\Teff=3600\K$, $\logg=4.0$, $\mh=-0.5$, $\alpha=1.5$ (middle panel);
$\Teff=3700\K$, $\logg=4.5$, $\mh=-0.0$, $\alpha=2.0$ (bottom panel).
}
\end{figure}

\begin{figure}
\psfig{file=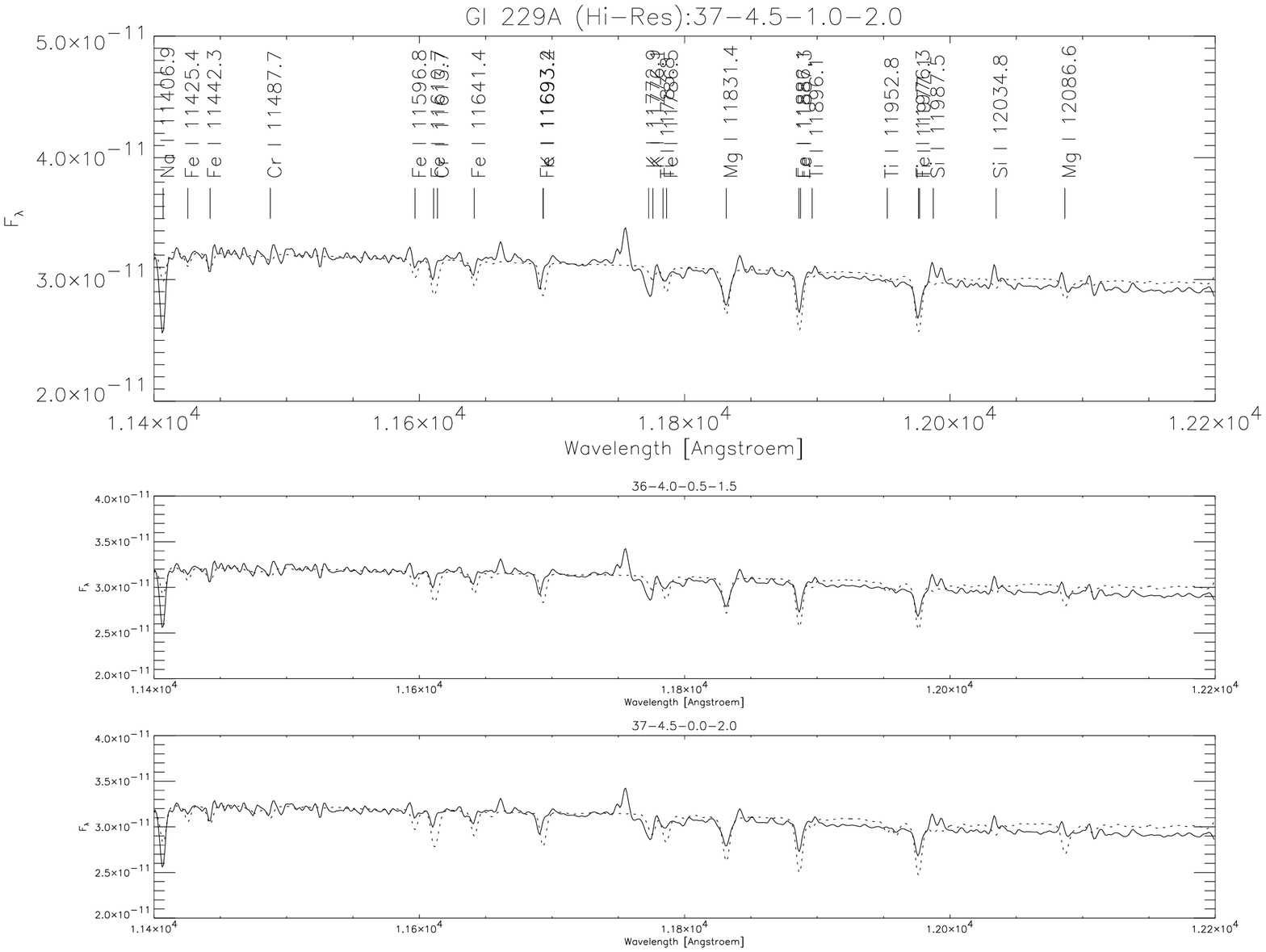,width=0.99\hsize,angle=90}
\caption[]{\label{gl229a-hr-long}Fit to higher resolution spectrum of Gl 229A using
best fit high-res spectra plus a number of close-by models. The parameters of the models (dotted lines) are 
$\Teff=3700\K$, $\logg=4.5$, $\mh=-1.0$, $\alpha=2.0$ (top panel);
$\Teff=3600\K$, $\logg=4.0$, $\mh=-0.5$, $\alpha=1.5$ (middle panel);
$\Teff=3700\K$, $\logg=4.5$, $\mh=-0.0$, $\alpha=2.0$ (bottom panel).
}
\end{figure}

The comparison of the Ames grid models to the high resolution spectra is shown in 
Figs.~\ref{gl229a-hr-short} and \ref{gl229a-hr-long}. Fig.~\ref{gl229a-hr-short}
indicates that scaled solar metallicities may not be a good assumption for Gl
229A: the lines of Fe~I are systematically too strong whereas the lines of Na~I are
too weak in the synthetic spectra. The K~I $\lambda1177$ doublet is too
weak in the models (Fig.~\ref{gl229a-hr-long}), but this comparison is uncertain 
because just longward of the K~I doublet there is some blending with Fe~I lines. 
We note that \cite{sch97} derived [Fe/H]$=-0.2$ for Gl~229A based on the FeH band at
1$\mu$m, however \cite{mou78} found $\mh = +0.15$ based on  Fourier analysis of 
Al, Ca and Mg infrared lines.  These conflicting results in the literature may
support a  non--solar metallicity pattern for this star.

\subsection{Gl 229B}

\begin{figure}
\psfig{file=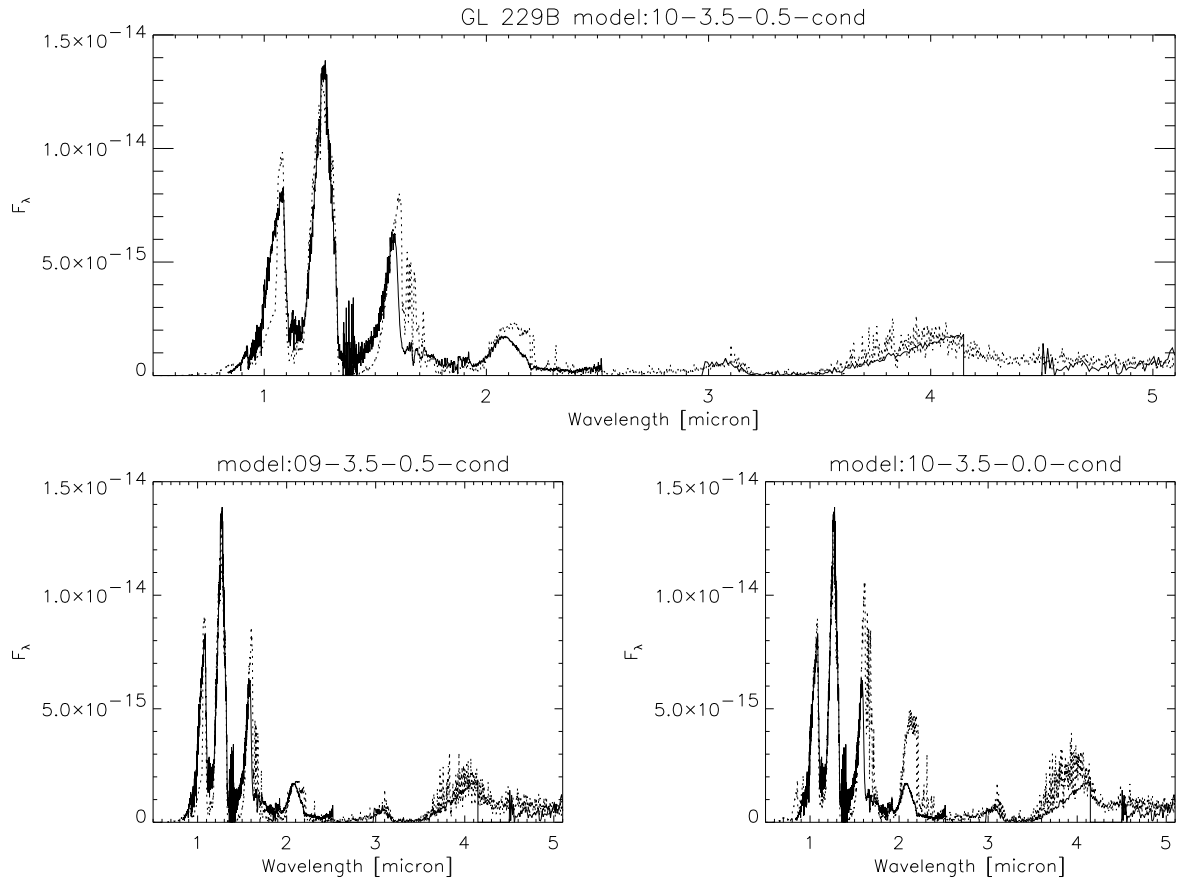,width=0.99\hsize,angle=90}
\caption[]{\label{gl229b}Best fit to Gl229B using the newest model grids.
The parameters of the models (dotted lines) are 
$\Teff=1000\K$, $\logg=3.5$, $\mh=-0.5$, (top panel);
$\Teff=900\K$, $\logg=3.5$, $\mh=-0.5$, (bottom left panel);
$\Teff=1000\K$, $\logg=3.5$, $\mh=0.0$, (bottom right panel).
}
\end{figure}

The best fits for Gl 229B are shown in Fig.~\ref{gl229b}.  For this object all 
model fitting was done by hand as the strong water and methane absorption bands
gave rise to erroneous results using the automated routine.  The range of
parameters for Gl 229B is $\Teff=1000\K\pm 100\K$, $\logg\la 3.5$ and
$\mh \approx -0.5$, putting  it firmly into the 
``AMES-cond'' regime; ``AMES-dusty'' models cannot fit its spectrum at all.  The results
corroborate the fits of Gl 229A with lower metallicity models so that the Gl
229 system does indeed seem to have low metallicities despite it being a
member of the kinematic young disk. Solar abundance models do in
general deliver worse fits than those with $\mh=-0.5$ (see Fig.~\ref{gl229b}), 
and models with even lower metallicities did not give better fits.
Other authors have also suggested that Gl~229B is metal-poor; \cite{gri00} find 
[O/H]$=-0.5$ based on their model analysis of the 0.8---1.0~$\mu$m spectrum, and 
\cite{sa00}, in an analysis of the same infrared spectra
studied here, find atmospheric parameters  $\Teff=870-1030\K$, $\logg =4.5-5.5$ and
$\mh =-0.5- -0.1$.  Although young metal--poor stars are uncommon they do exist,
as shown by \cite{fel01}, in particular their Figures 13(e) and 18(a).

\begin{figure}
\psfig{file=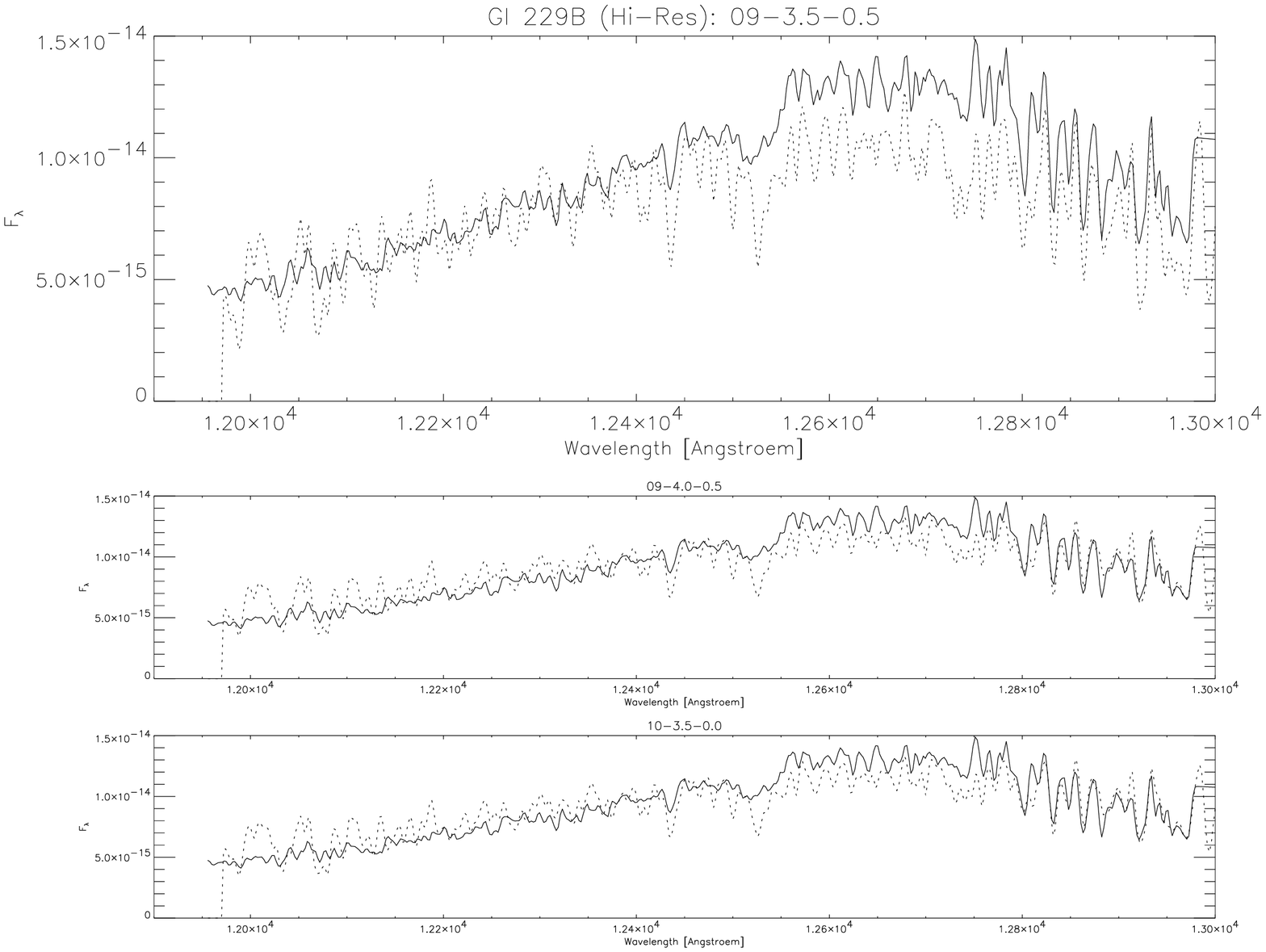,width=0.99\hsize,angle=90}
\caption[]{\label{gl229b-hr-short}Fit to higher resolution spectrum of Gl 229B using
best fit high-resolution spectrum plus a number of close-by models. The parameters of the models (dotted lines) are 
$\Teff=900\K$, $\logg=3.5$, $\mh=-0.5$ (top panel);
$\Teff=900\K$, $\logg=4.0$, $\mh=-0.5$ (middle panel);
$\Teff=1000\K$, $\logg=3.5$, $\mh=-0.0$ (bottom panel).
}
\end{figure}

\begin{figure}
\psfig{file=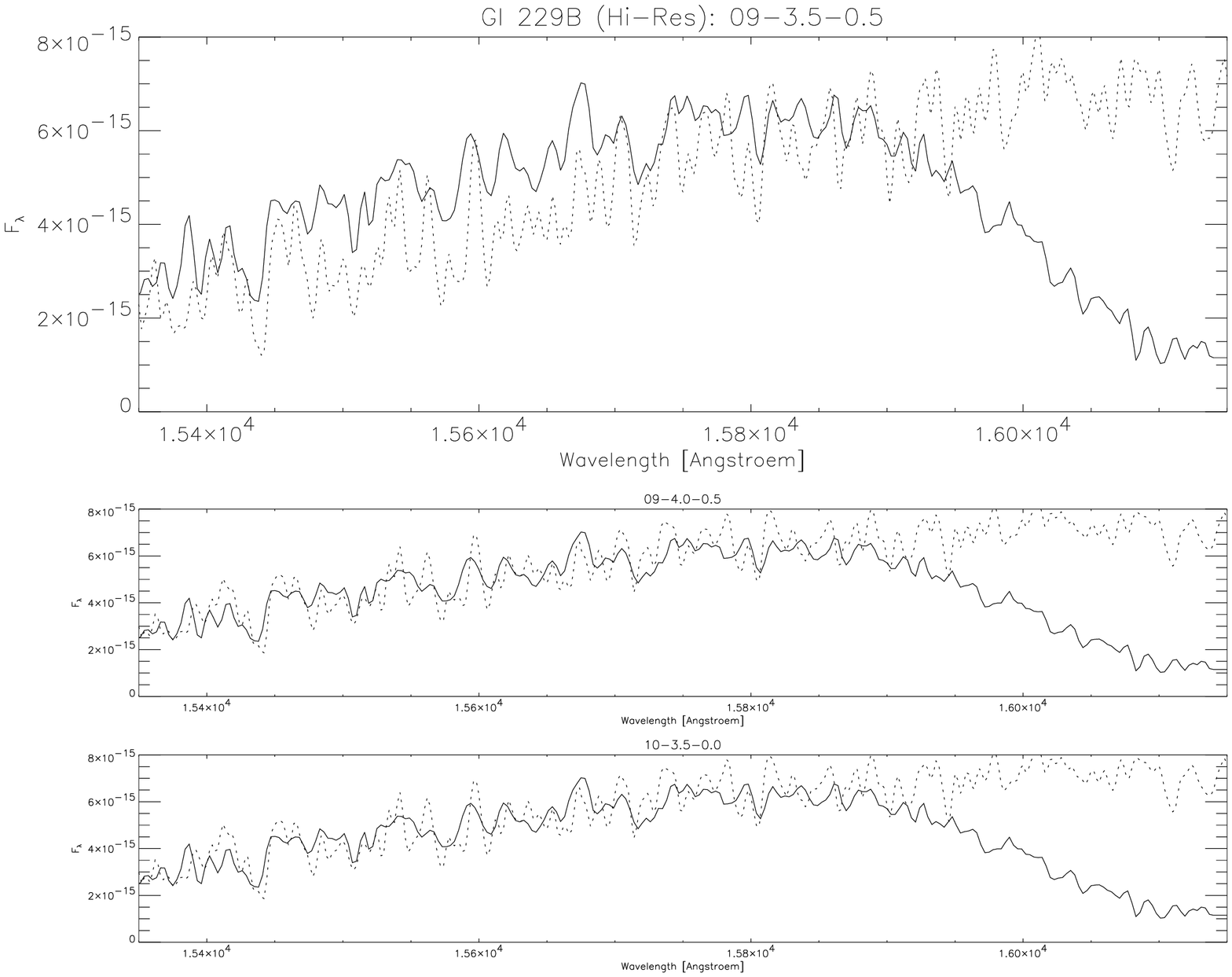,width=0.99\hsize,angle=90}
\caption[]{\label{gl229b-hr-mid}Fit to higher resolution spectrum of Gl 229B using
best fit high-resolution spectrum plus a number of close-by models. The parameters of the models (dotted lines) are 
$\Teff=900\K$, $\logg=3.5$, $\mh=-0.5$ (top panel);
$\Teff=900\K$, $\logg=4.0$, $\mh=-0.5$ (middle panel);
$\Teff=1000\K$, $\logg=3.5$, $\mh=-0.0$ (bottom panel).
}
\end{figure}

\begin{figure}
\psfig{file=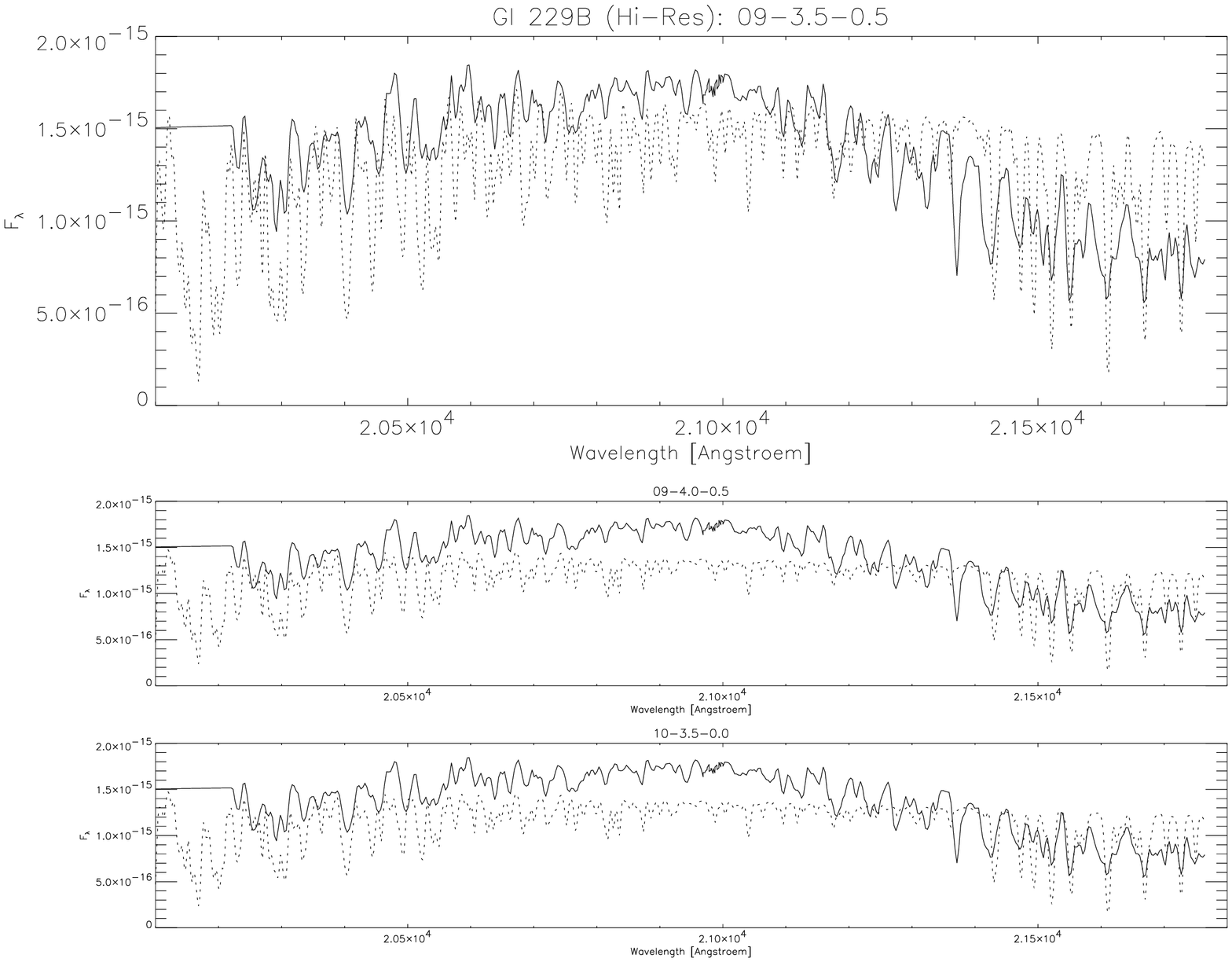,width=0.99\hsize,angle=90}
\caption[]{\label{gl229b-hr-long}Fit to higher resolution spectrum of Gl 229B using
best fit high-resolution spectrum plus a number of close-by models. The parameters of the models (dotted lines) are 
$\Teff=900\K$, $\logg=3.5$, $\mh=-0.5$ (top panel);
$\Teff=900\K$, $\logg=4.0$, $\mh=-0.5$ (middle panel);
$\Teff=1000\K$, $\logg=3.5$, $\mh=-0.0$ (bottom panel).
}
\end{figure}

The fits to high resolution spectra of Gl 229B are shown in
Figs.~\ref{gl229b-hr-short}--\ref{gl229b-hr-long}.  For the most part the synthetic 
spectra fit the atomic line profiles quite well, but significant discrepancies 
remain, in particular in the $1.5$--$1.7\,\mu$m range (Fig.~\ref{gl229b-hr-mid}). 
These problems are likely caused in part by incomplete molecular line data, in
particular water vapor and methane. The analysis roughly confirms the results from
the low resolution spectra.

\section{Comparison to Evolutionary Models}

Table 5 lists the best fit model parameters, together with the observed luminosities 
of each of our targets.  These results can be used to constrain mass and age for 
each component through evolutionary models.   We used recent evolutionary 
models but were obliged to use three different models for the comparison as each
covered a limited range in temperature and mass.  For the M dwarfs LHS 102A and Gl 229A 
we used the solar-metallicity and metal-poor models of \cite{1998A&A...337..403B}, 
for the L dwarf LHS 102B those by \cite{2000ApJ...542..464C}, and for the T dwarf 
Gl 229B those by \cite{bur97,bur01}.  
The first two models use earlier generation ``AMES-dusty'' atmospheres, the last a 
``cond''--type atmosphere with the dust not contributing to the atmospheric 
opacity.  The T dwarf comparison is  based on a solar metallicity evolutionary 
model and is therefore more uncertain than the others.  We constrain the age of
Gl 229B to be that of Gl 229A, and use the evolutionary models to determine mass
based on age, luminosity, temperature and gravity.  These give self-consistent 
results, however \cite{bur01} show that for a given luminosity or temperature, and 
fixed age, a more metal-poor brown dwarf has a higher mass than a solar metallicity 
brown dwarf, and so we regard the value derived here for the mass of Gl~229B to be 
a lower limit.

To determine the possible ranges in mass and age for each component of the two 
binaries, we adopted uncertainties in effective temperature of 100~K, in $\logg$ 
of 0.5~dex, and in log$_{10}L/L_{\odot}$ of 0.02~dex.  The metallicity was 
considered to be fixed as evolutionary models are only available with either 
$\mh =$0 or $\mh =-$0.5.  Assuming that the components of each binary have the 
same age, we find that the Gl~229 system is very young (aged 16---45~Myr) with 
masses for (A,B) of (0.30---0.45,$\ga$0.007)M$_{\sun}$, and the LHS~102 system 
is older, aged 1---10~Gyr with masses (0.19,0.07)M$_{\sun}$.  The ages are 
consistent with the kinematically implied disk populations (Table 1), although 
\cite{nak95} suggests that the lack of H$\alpha$ emission in Gl~229A implies an 
age older than 100~Myr.  We note however that the star is a flare star, which 
supports a young age.  The young age of the Gl~229 system gives rise to a lower 
mass for the T dwarf than previously assigned, only about 7 Jupiter masses, 
although as mentioned above this is a lower limit.  The B component of the LHS 102 
system is most likely a brown dwarf, i.e. just on the substellar side 
of the the stellar/substellar mass boundary, based on its luminosity 
\cite[see for example Figure 1 of][]{bur01}.

\section{Conclusions}

In this paper we have presented the results of the analysis of two binary 
systems, one (LHS 102) consisting of an M/L dwarf pair and one (Gl 229) 
being an M/T dwarf system. 
The M dwarf primaries show that the best mixing length for effective temperatures
above $3000\K$ is somewhere around $1.5$ to $2.0$, which is consistent with recent
hydrodynamical models \cite[]{ludwig-pap}.

The LHS 102 system is  best fit with solar abundance models.  LHS 102B is an 
L dwarf with an effective temperature around $1900\K$ that is best fit with 
``AMES-dusty'' models that include the effects of dust formation and opacity on the 
atmospheric structure and the emitted spectrum. More work is required on
the detailed treatment of photospheric dust to produce a good match to
the entire observed flux distribution; such work is in progress.
LHS 102B is found to be just below the stellar mass limit.

The results for Gl 229 indicate that this system is metal deficient with 
$\mh \approx -0.5$.  We have determined an age for the Gl 229 system of
about 30 Myr, constrained primarily by the observed luminosity 
and derived effective temperature of the A component.  This young age
is consistent with the low surface gravity derived for Gl 229B
\cite[see for example Figure 9 of][]{bur97}, and translates to 
a very low mass for this T dwarf of $\ga$~7 Jupiter masses. A better mass
determination requires metal-poor evolutionary models of brown dwarfs, which 
are not currently available.

\section*{Acknowledgments}

We are very grateful to the staff at UKIRT for their assistance in obtaining
the data presented in this paper.  Some of the data presented here were
obtained through the UKIRT Service Programme.  UKIRT, the United Kingdom
Infrared Telescope, is operated by the Joint Astronomy Centre on behalf of the
U.K. Particle Physics and Astronomy Research Council. TRG is supported by the 
Gemini Observatory, which is operated by the Association of Universities for 
Research in Astronomy, Inc., on behalf of the international Gemini parthership 
of Argentina, Australia, Brazil, Canada, Chile, the United Kingdom and the 
United States of America. FA acknowledges support from  NASA LTSA NAG5-3435 
and NASA EPSCoR grants to Wichita State University, and support from CNRS.
This work was supported in part by NSF grants AST-9720704 and AST-0086246, 
NASA grants NAG5-8425, NAG5-9222, as well as NASA/JPL grant 961582 to the 
University of Georgia and in part by NSF grants AST-97314508, by NASA grant 
NAG5-3505 and an IBM SUR grant to the University of Oklahoma.  This work also 
was supported in part by the P\^ole Scientifique de Mod\'elisation Num\'erique 
at ENS-Lyon.  Some of the calculations presented in this paper were performed 
on the IBM SP2 of the UGA UCNS, on the IBM SP ``Blue Horizon'' of the San Diego
Supercomputer Center (SDSC), with support from the National Science Foundation,
and on the IBM SP of the NERSC with support from the DoE.  We thank all these 
institutions for a generous allocation of computer time.

\newpage

\bibliography{leggett}

\newpage

\begin{table}
\caption{Sample Description}
\begin{tabular}{rrclc}
Name & Other & RA/Dec & Spectral & Kinematic  \\
  & Names  & equinox 2000/epoch 1999 & Type & Population \\
\hline
LHS 102A         &  GJ 1001   &   00:04:36.37 $-$40:44:02.5 & dM3.5 & old disk \\
LHS 102B         &            &   00:04:34.83 $-$40:44:06.0 & dL4.5 & \\
Gliese 229A      &  LHS 1827  &   06:10:34.69 $-$21:51:48.9 & dM1   & young disk \\
Gliese 229B      &            &   06:10:34.86 $-$21:51:56.3 & dT6   & \\
\end{tabular}
\end{table}

\begin{table}
\caption{New Photometry}
\begin{tabular}{rrrrrrr}
Name  & $I$ (error) & $Z$ (error) & $J$  (error) & $H$  (error) & $K$ (error) & $L^{\prime}$ (error) \\
& \multicolumn{2}{c}{(UFTI)} & \multicolumn{4}{c}{(MKO)}  \\
\hline
LHS 102A &  9.55 (0.03) &  9.12 (0.03) &  8.62 (0.03) &  8.02 (0.03) &  7.73 (0.03) &  7.53 (0.05) \\
LHS 102B & 15.89 (0.04) & 14.67 (0.03) & 13.06 (0.03) & 12.14 (0.03) & 11.36 (0.03) & 10.41 (0.05) \\
Gl 229A  & \ldots      & \ldots      &  \ldots     & \ldots      & \ldots      &   4.06 (0.05) \\
\end{tabular}
\end{table}

\begin{table}
\caption{Colours and Fluxes for the Sample }
\begin{tabular}{rrrrrrrrrrrrr}
Name & $M-m$ &  $I_C$ & $I_C -Z$ &  $I_C -J$ & $J-H$
 & $H-K$ & $K$ & $K-L^{\prime}$ & Flux & m$_{bol}^a$ & BC$_{K}^b$ &
log $_{10}$ \\
 &    &  & & \multicolumn{5}{c}{(MKO)} &  W/m$^2$ & & & L/L$_{\odot}^c$ \\
\hline
LHS 102A & 0.0997 &  9.86 & 0.74 & 1.24 &    0.60 &    0.29 &  7.73 & 0.20 & 1.69e$-$12 & 10.45 &  2.72 & $-$2.32 \\
LHS 102B & 0.0997 & 16.68 & 2.01 & 3.63 &    0.92 &    0.78 & 11.36 & 0.95 & 3.12e$-$14 & 14.79 &  3.43 & $-$4.05 \\
Gl 229A  & 1.1926 &  6.11 & 0.43 & 1.05 &    0.65 &    0.25 &  4.15 & 0.09 & 4.98e$-$11 &  6.78 &  2.63 & $-$1.29 \\
Gl 229B  & 1.1926 & 20.02 & 3.84 & 6.01 & $-$0.35 &    0.00 & 14.36 & 2.14 & 5.97e$-$15 & 16.58 &  2.22 & $-$5.21 \\
\\
\end{tabular}
Notes\\
$^a$: Adopting L$_{\odot} = 3.86e26$ W and M$_{\rm bol, \odot} = 4.75$ then:
$ m_{bol} = -2.5\times log_{10}(flux) - 18.978 $ \\
$^b$: BC$_K = m_{bol} -$ K\\
$^c$: Adopting L$_{\odot} = 3.86e26$ W, and if 
$\pi$ is parallax in arcseconds then:
$ log_{10}L/L_{\odot} = log_{10}(flux) -2\times log_{10}\pi + 7.491 $\\
\end{table}

\begin{table}
\caption{Atmospheric Model Fit Parameters}
\begin{tabular}{rrr}
Name  & $T_{\rm eff}$/log(g)/[m/H] & Quality \\
\hline
LHS 102A & 3200/5.0/0.0  &  7.92 \\
LHS 102A & 3200/4.5/0.0  &  7.94 \\
LHS 102A & 3200/5.5/0.0  &  7.98 \\
LHS 102A & 3200/5.0/-0.5  &  8.12 \\
LHS 102A & 3000/5.0/-0.7  &  8.29 \\
LHS 102B & 1900/6.0/0.0  & 15.68 \\
LHS 102B & 1900/5.5/0.0  & 16.21 \\
LHS 102B & 2000/5.5/0.0  & 16.50 \\
Gl 229A  & 3700/4.0/-0.5  &  7.44  \\
Gl 229A  & 3700/3.5/-0.7  &  7.44  \\
Gl 229A  & 3600/3.5/-1.0  &  7.47  \\
Gl 229A  & 3500/4.5/0.0  &  9.87 \\
Gl 229A  & 3400/5.5/0.0  & 10.08 \\
Gl 229B  & 1000/3.5/-0.5  & \ldots   \\
\end{tabular}
\end{table}

\begin{table}
\caption{Evolutionary Parameters}
\begin{tabular}{rrrrrcc}
Name  & L/L$_{\odot}$ & [m/H] & $T_{\rm eff}$ K & log(g) & Mass M$_{\sun}$& Age Gyr  \\
\hline
LHS 102A & -2.32 &  0.0 & 3200 & 5.0 & 0.13---0.19 & 0.08---10.0\\
LHS 102B & -4.05 &  0.0 & 1900 & 6.0 & 0.065---0.074 & 1---10 \\
Gl 229A  & -1.29 & -0.5 & 3700 & 4.0 & 0.30---0.45 & 0.016---0.045\\
Gl 229B  & -5.21 & -0.5 & 1000 & 3.5 & $\ga$0.007 & \ldots \\
\end{tabular}
\end{table}

\newpage

\end{document}